\documentclass[pra,aps,amsfonts,amsmath,twocolumn,showpacs,floatfix]{revtex4}
\usepackage{bm}
\usepackage{amsfonts}
\usepackage{amsmath}
\usepackage{graphicx}

\newcommand{\Eq}[1]{Eq.~(\ref{#1})}
\newcommand{\Eqs}[1]{Eqs.~(\ref{#1})}
\newcommand{\Ref}[1]{Ref.~\cite{#1}}
\newcommand{\Fig}[1]{Fig.~\ref{#1}}

\newcommand{\Sec}[1]{Sec.~\ref{#1}}
\DeclareMathOperator{\Real}{Re}

\renewcommand{\Re}{\Real}

\newcommand{\poisson}[2]{\{#1,#2\}}

\newcommand{\deltat}{\tilde{\delta}}
\newcommand{\Deltat}{\tilde{\Delta}}

\newcommand{\fhat}{\hat{f}}
\newcommand{\fp}{f^\prime}
\newcommand{\htil}{\tilde{h}}
\newcommand{\jv}{\bm{\mathrm{j}}}
\newcommand{\kappat}{\tilde{\kappa}}
\newcommand{\Lv}{\bm{\mathrm{L}}}
\newcommand{\nablav}{\bm{\nabla}}
\newcommand{\nud}{\dot{\nu}}
\newcommand{\phid}{\dot{\phi}}
\newcommand{\phidd}{\ddot{\phi}}
\newcommand{\pv}{\bm{\mathrm{p}}}
\newcommand{\pvh}{\bm{\mathrm{\hat{p}}}}
\newcommand{\Pv}{\bm{\mathrm{P}}}
\newcommand{\Pvd}{\dot{\bm{\mathrm{P}}}}
\newcommand{\rhod}{\dot{\rho}}
\newcommand{\rv}{\bm{\mathrm{r}}}
\newcommand{\Rv}{\bm{\mathrm{R}}}
\newcommand{\Rvd}{\dot{\bm{\mathrm{R}}}}
\newcommand{\varrhot}{\tilde{\varrho}}
\newcommand{\varrhotd}{\dot{\tilde{\varrho}}}
\newcommand{\vhat}{\hat{v}}
\newcommand{\vv}{\bm{\mathrm{v}}}
\newcommand{\Vhat}{\hat{V}}

\newcommand{\icoll}{\mathit{coll}}
\newcommand{\iev}{\mathit{ev}}
\newcommand{\iod}{\mathit{od}}
\newcommand{\iext}{\mathit{ext}}
\newcommand{\iHartree}{\mathit{Hartree}}

\newcommand{\imax}{\mathit{max}}

\begin{document}
\title{Coupling of hydrodynamics and quasiparticle motion in
  collective modes of superfluid trapped Fermi gases}
\author{Michael Urban}
\affiliation{Institut de Physique Nucl{\'e}aire, CNRS and
  Univ.~Paris-Sud, 91406 Orsay C{\'e}dex, France}
\begin{abstract}
At finite temperature, the hydrodynamic collective modes of superfluid
trapped Fermi gases are coupled to the motion of the normal component,
which in the BCS limit behaves like a collisionless normal Fermi
gas. The coupling between the superfluid and the normal components is
treated in the framework of a semiclassical transport theory for the
quasiparticle-distribution function, combined with a hydrodynamic
equation for the collective motion of the superfluid component. We
develop a numerical test-particle method for solving these equations
in the linear response regime. As a first application we study the
temperature dependence of the collective quadrupole mode of a Fermi
gas in a spherical trap. The coupling between the superfluid
collective motion and the quasiparticles leads to a rather strong
damping of the hydrodynamic mode already at very low temperatures. At
higher temperatures the spectrum has a two-peak structure, the second
peak corresponding to the quadrupole mode in the normal phase.
\end{abstract}
\maketitle
%
\section{Introduction}
Most of the current experiments involving trapped atomic Fermi gases
focus on the BEC-BCS crossover. By changing the magnetic field around
a Feshbach resonance, the scattering length $a$ of the atoms can be
varied from small positive values through very large values near the
resonance to small negative values. For $a>0$, $k_F a\ll 1$ (where
$k_F$ denotes the Fermi momentum) the system can be considered as a
Bose-Einstein condensate (BEC) of diatomic molecules. The crossover
region, $k_F |a| \gtrsim 1$, is not yet very well understood from a
theoretical point of view. Finally, on the other side of the
resonance, when $a<0$, $k_F|a|\ll 1$, the system should be in the BCS
phase if the temperature is sufficiently low. However, the BCS
critical temperature $T_c$ is extremely low, and very soon the
magnetic field reaches the point where $T_c$ becomes smaller than the
actual temperature $T$, and the system undergoes the phase transition
to the normal (non-superfluid) phase.

One possibility to study the crossover experimentally is to measure
the properties of certain collective oscillations. For example, the
radial and axial breathing modes of a cigar-shaped trapped Fermi gas
have been measured over the whole crossover region
\cite{Bartenstein,Kinast}. In these experiments one can observe how
the frequencies and damping rates of the modes change from what one
expects for a BEC to what one expects for a collisionless normal Fermi
gas. Assuming that, except in the collisionless normal phase,
hydrodynamics is valid, the measured frequencies can give some
information on the equation of state in the crossover region.

However, this schematic picture is not completely accurate. Since the
system is in a trap, there is no sharp transition from the superfluid
to the normal phase. This can be seen as follows: The BCS critical
temperature $T_c$ depends on the atom density $\rho$, and the density
depends on the position $\rv$. In the center of the trap, the density
$\rho(\rv)$ and hence the local critical temperature $T_c(\rv)$ are
higher than in the outer part of the trap. As a consequence, for a
given temperature, the outer part gets already normal at a magnetic
field where the inner part is still superfluid. To be more precise, a
system in the BCS phase at finite temperature behaves effectively like
a mixture of superfluid and normal components with densities $\rho_s$
and $\rho_n$, respectively, which become $\rho_s = \rho$, $\rho_n=0$
in the limit $T = 0$ and $\rho_s=0$, $\rho_n = \rho$ in the limit $T
\geq T_c$. As a consequence, if $0< T < T_c(\rv=0)$, the superfluid
inner part of the trap behaves like a mixture of normal and superfluid
components, while only the outer part with $T_c(\rv) < T$ is
completely normal \cite{Urban}.

If the collision rate was high enough, also the normal component of
the gas would behave hydrodynamically. Such a system could be
described by Landau's two-fluid hydrodynamics which has been applied
to collective modes in trapped superfluid gases at finite temperature
\cite{TaylorGriffin}. However, although in the recent experiments the
transition to the normal phase seemed to occur at a value of $k_F |a|
\approx 2$ \cite{Bartenstein} (i.e., the BCS phase has not really been
reached), the system behaved already like a collisionless normal Fermi
gas. Hence it seems to be clear that the normal component cannot be
treated in terms of hydrodynamics, but a description in terms of a
Vlasov equation is required.

We note that there are other approaches to the description of the
collective modes at finite temperature. In particular, let us mention
the quasiparticle random phase approximation (QRPA)
\cite{BruunQRPA,GrassoKhan}, which can be seen as the linearized form
of the time-dependent Bogoliubov-de Gennes (BdG) equations. However,
for practical reasons this method is limited to systems with spherical
symmetry and numbers of particles up to a few times $10^4$. Another
disadvantage of this method is that it does not allow to include a
collision term.

For the case of clean superconductors, a semiclassical transport
theory taking the coupling between normal and superconducting
components into account has been developed by Betbeder-Matibet and
Nozi\`eres \cite{BetbederMatibet}. Transport theories of this type
have also been used for describing the dynamics of superfluid $^3$He
\cite{SereneRainer,VollhardtWoelfle}. In a preceding paper
\cite{UrbanSchuck}, we derived the semiclassical transport equations
for the case of trapped atomic Fermi gases and applied them to the
quadrupole mode of a gas in a spherical trap. We found that the
presence of the normal component leads to a strong damping of the
hydrodynamic collective mode. The same mechanism might explain the
strong damping observed experimentally near the transition to the
collisionless behavior \cite{Bartenstein}. However, in
\Ref{UrbanSchuck} we had to replace the gap $\Delta(\rv)$ by a
constant in order to find an analytical solution of the transport
equations. Due to this simplification, which cannot really be
justified, the damping of the hydrodynamic mode at a given temperature
was much weaker than that obtained in QRPA calculations
\cite{GrassoKhan}.

In the present paper we will work out a numerical method which allows
us to treat the realistic $\rv$-dependence of the gap. In addition,
the method is very versatile and allows to treat much more general
cases than can be solved analytically in the constant-gap
approximation. The basic idea is to replace the continuous phase-space
distribution function of the quasiparticles by a sum of a finite
number of delta functions in phase space, called ``test particles.''
In the normal phase, the test-particle method is routinely used for
solving the Vlasov equation, e.g. for simulating heavy-ion collisions
in nuclear physics \cite{Bertsch}. It has also been applied to the
simulation of the dynamics of normal trapped atomic Fermi gases with
collision term \cite{Toschi} and of Bose-Fermi mixtures
\cite{Maruyama}. However, to our knowledge, the test-particle method
has not yet been used in the context of superfluid systems, and in
fact the numerical difficulties are quite different from those
encountered in the usual applications.

The article is organized as follows. In \Sec{transportbcs}, we give a
brief summary of the transport equations for the BCS phase and their
linearization in the case of small deviations from equilibrium. We
also give arguments why some terms which appear in the equations can
be neglected. In \Sec{testparticlemethod} we introduce the
test-particle method for the case of small oscillations around
equilibrium. We describe in detail a number of tricky points we
encountered during the implementation of the method, in particular the
calculation of the test-particle trajectories, the generation of the
test-particle distribution in phase space, and the initialization
after a delta-like perturbation. In \Sec{firstresults} we present the
first results obtained with the help of this method, again for the
quadrupole mode in a spherical system. Finally, in \Sec{conclusions},
we summarize and draw our conclusions.
\section{Transport equations for the BCS phase}
\label{transportbcs}

\subsection{Summary of the kinetic equations}
\label{kineticsummary}
In this subsection we will give a brief summary of the kinetic
equation approach developed by Betbeder-Matibet and Nozi\`eres
\cite{BetbederMatibet} for the case of clean superconductors and
adapted to the case of trapped atomic Fermi gases in
\Ref{UrbanSchuck}. In this paper we will only give the final
equations. For details of the derivations, see \Ref{UrbanSchuck} and
the footnote
\footnote{There are some typos in \Ref{UrbanSchuck} which we wish to
correct here: In Eq.~(33), $\Delta$ should be replaced by
$\Deltat$. In Eq.~(34b), $\varrhotd_\iev$ should be replaced by
$\nud_\iev$. In Eq.~(61), $\Delta_{1\nu}(\rv)$ should be replaced by
$-\Delta_{1\nu}(\rv)$. In addition, we point out that $\Deltat_1 = 0$
if $\Delta_0 = 0$, which is not evident from Eq.~(61) but can be
derived from Eq.~(54).}.

We consider a dilute gas of fermionic atoms of mass $m$ in two equally
populated hyperfine states $\uparrow$ and $\downarrow$, trapped by an
external potential $V_\iext$ and interacting via an attractive
short-range interaction which leads to a scattering length $a <
0$. The corresponding classical mean-field hamiltonian (minus the
chemical potential $\mu$) reads
\begin{equation}
h(\rv,\pv) = \frac{\pv^2}{2m}+V(\rv)-\mu\,,
\end{equation}
where $V$ denotes the sum of the external and the Hartree potential,
\begin{equation}
V(\rv) = V_\iext(\rv)+V_\iHartree(\rv) = V_\iext(\rv)+g\rho(\rv)\,.
\end{equation}
In the latter equation, $g = 4\pi\hbar^2 a/m$ denotes the coupling
constant and $\rho$ is the density per spin state. The Vlasov equation
(without collision term) for the phase-space distribution function
$\varrho(\rv,\pv)$ in the normal phase can be written in the compact
form
\begin{equation}
\dot{\varrho} = \{h,\varrho\}\,, 
\label{vlasov}
\end{equation}
where $\{\cdot,\cdot\}$ denotes the Poisson bracket. One way to derive
this equation is to perform a Wigner-Kirkwood expansion up to order
$\hbar$ of the time-dependent Hartree-Fock equation
\cite{Bertsch,RingSchuck}.

In the superfluid phase the derivation of an analogous transport
equation is much more complicated due to the presence of the complex
order parameter (gap) $\Delta(\rv)$ whose phase describes the
collective motion of the Cooper pairs. In addition to the density
matrix $\varrho$, there exists now an anomalous density matrix
(pairing tensor) $\kappa$. The gap $\Delta$ and the anomalous density
are related by the gap equation
\begin{equation}
\Delta(\rv) = -g \int \frac{d^3p}{(2\pi\hbar)^3}
  \Big(\kappa(\rv,\pv)-\frac{\Delta(\rv)}{p^2/m}\Big)\,.
\end{equation}
The time-dependence of $\varrho$ and $\kappa$ is governed by the
time-dependent Hartree-Fock-Bogoliubov or BdG equations. As in the
normal phase, the semiclassical transport theory can be derived from
these equations by performing a Wigner-Kirkwood expansion up to order
$\hbar$. However, it turns out that it is necessary to introduce a
gauge transformation with a phase $\phi(\rv)$ that makes the order
parameter $\Delta$ real. This corresponds to a transformation into the
local rest frame of the Cooper pairs, which are moving with the
collective velocity $\vv_\icoll(\rv) =
-(\hbar/m)\nablav\phi(\rv)$. The effect of this transformation is to
change the gap $\Delta$, the single-particle hamiltonian $h$, the
normal and anomalous density matrices $\varrho$ and $\kappa$ according
to
\begin{align}
\Deltat(\rv) &= \Delta(\rv) e^{2i\phi(\rv)}\equiv
  |\Delta(\rv)|\,,\\
\htil(\rv,\pv) &=
h[\rv,\pv-\hbar\nablav\phi(\rv)]-\hbar\phid(\rv)\,,\label{htil}\\ 
\varrhot(\rv,\pv) &= \varrho[\rv,\pv-\hbar\nablav\phi(\rv)]\,,\\
\kappat(\rv,\pv) &= \kappa(\rv,\pv)e^{2i\phi(\rv)}\,.
\end{align}

Roughly speaking, the phase $\phi$ determines the dynamics of the
superfluid component of the system, while the dynamics of the normal
component, consisting of thermally excited quasiparticles, has to be
described separately. The distribution of these quasiparticles,
denoted by $\nu(\rv,\pv)$, obeys the following equation of motion:
\begin{equation}
\nud = \poisson{E}{\nu}\,.
\label{qpvlasov}
\end{equation}
This equation looks formally very similar to the Vlasov equation
(\ref{vlasov}), except that the hamiltonian $h$ is replaced by the
quasiparticle energie $E$, which is defined as
\begin{equation}
E = \sqrt{\htil_\iev^2+\Deltat^2}+\htil_\iod\,.
\end{equation}
Throughout this article, the indices ``$\iev$'' and ``$\iod$'' denote the
time-even and time-odd parts of a phase-space function, i.e., the
parts which are even and odd in $\pv$, respectively. The
quasiparticle-distribution function $\nu$ is related to the normal and
anomalous density matrices in the new gauge, $\varrhot$ and $\kappat$,
by
\begin{gather}
\varrhot = \frac{1}{2}-\frac{\htil_\iev}{2E_\iev}(1-2\nu_\iev)+\nu_\iod\,,\\
\Re \kappat = \frac{\Deltat}{2E_\iev}(1-2\nu_\iev)\,.\label{Rekappat}
\end{gather}

The Vlasov-like equation (\ref{qpvlasov}) has to be complemented with
an equation of motion for the phase $\phi$. It turns out that $\phi$
has to be determined from the continuity equation
\begin{equation}
\rhod(\rv)+\nablav\cdot\jv(\rv) = 0\,,
\label{conti}
\end{equation}
where the density $\rho$ and the current $\jv$ are given by
\begin{align}
\rho(\rv) &= \int \frac{d^3p}{(2\pi\hbar)^3}\varrhot(\rv,\pv)\,,\\
\jv(\rv) &= \int \frac{d^3p}{(2\pi\hbar)^3}
  \frac{\pv}{m}\,\varrhot(\rv,\pv)
  -\frac{\hbar}{m}\rho(\rv)\nablav\phi(\rv)\,.
\label{defcurrent}
\end{align}

\subsection{Linearization around equilibrium}
\label{linearization}
Let us now assume that the external potential $V_\iext$ can be written
as
\begin{equation}
V_\iext = V_{0\iext} + V_{1\iext}\,,
\end{equation}
where $V_{0\iext}$ is time-independent and $V_{1\iext}$ is a small
perturbation. The equilibrium quantities (corresponding to the
potential $V_{0\iext}$) will be marked by an index ``0''. In
particular, since in equilibrium the gap can be chosen to be real, we
have
\begin{equation}
\phi_0 = 0\,,\quad \htil_0 = h_0\,,\quad \Deltat_0 = \Delta_0\,.
\end{equation}
The quasiparticle distribution function in equilibrium is given by
\begin{equation}
\nu_0(\rv,\pv) = f[E_0(\rv,\pv)]\,,\quad 
\end{equation}
where $f(E)$ denotes the Fermi function,
\begin{equation}
f(E) = \frac{1}{e^{E/(k_B T)}+1}\,.
\end{equation}
Our aim is to calculate the small deviations from equilibrium induced
by the perturbation $V_{1\iext}$, which will be marked by an index
``1''. To that end we linearize the transport equation
(\ref{qpvlasov}) for the quasiparticle-distribution function,
\begin{equation}
\nud_1 - \poisson{E_0}{\nu_1} 
  = \fp(E_0)\poisson{E_1}{E_0}\,,
\label{qpvlasovlin1}
\end{equation}
where $\fp(E_0) = df/dE_0$. The deviation of the quasiparticle energy,
$E_1$, which appears on the r.h.s., depends itself on $\nu_1$ through
the deviation of the Hartree field, $g\rho_1$, and the deviation of
the gap, $\Deltat_1$. Expressing $\rho_1$ and $\Deltat_1$ in terms of
$\nu_1$, we can write \Eq{qpvlasovlin1} as
\begin{multline}
\nud_1-\poisson{E_0}{\nu_1} =\\
  -\frac{\fp(E_0)}{m}\Big(-\pv\cdot\nablav
    \frac{V_{1\iext}+g\rho_{1\nu}-\hbar\phid_1}{1+gA}\\
  +\frac{\Delta_0}{E_0^2}\pv\cdot\nablav
    \frac{\Delta_0(V_{1\iext}+g\rho_{1\nu}-\hbar\phid_1)}{1+gA}\\
  +\frac{h_0}{E_0^2}\pv\cdot\nablav
    \frac{\Delta_0\Delta_{1\nu}}{gA}
  +\frac{\hbar}{m}\frac{h_0}{E_0}(\pv\cdot\nablav)^2\phi_1\\
  -\hbar\frac{h_0}{E_0}(\nablav V_0)\cdot\nablav\phi_1
  -\hbar\frac{\Delta_0}{E_0}(\nablav\Delta_0)\cdot\nablav\phi_1\Big)\,.
\label{qpvlasovlin}
\end{multline}
where $\rho_{1\nu}$ and $\Delta_{1\nu}$ are the quasiparticle
contributions to $\rho_1$ and $\Deltat_1$,
\begin{gather}
\rho_{1\nu}(\rv) = \int\frac{d^3p}{(2\pi\hbar)^3}
  \frac{h_0(\rv,\pv)}{E_0(\rv,\pv)}\nu_1(\rv,\pv)\,,\\
\Delta_{1\nu}(\rv) = g\int\frac{d^3p}{(2\pi\hbar)^3}
  \frac{\Delta_0(\rv)}{E_0(\rv,\pv)}\nu_1(\rv,\pv)\,,
\end{gather}
while $A(\rv)$ is a function which depends only on equilibrium
quantities. The explicit expression for the function $A(\rv)$ reads
\begin{equation}
A(\rv) = \frac{mp_F(\rv)}{2\pi^2\hbar^3}[1-\varphi(\rv)]\,,
\end{equation}
where the local Fermi momentum $p_F(\rv)$ is defined as usual by
$p_F^2(\rv)/(2m) = \epsilon_F(\rv) = \mu-V_0(\rv)$, and the
temperature dependence of $A(\rv)$ is governed by the function
\begin{equation}
\varphi(\rv) = -\int d\xi\frac{\xi^2}{E_\xi^2}\fp(E_\xi)\,,
\label{defphi}
\end{equation}
with $E_\xi = \sqrt{\xi^2+\Delta_0^2(\rv)}$. In the two limiting cases
$T = 0$ and $T \geq T_c(\rv)$, the function $\varphi(\rv)$ takes the
values 0 and 1, respectively. As a consequence, $A(\rv) = 0$ if $T
\geq T_c(\rv)$.

In order to determine the phase $\phi_1$, we also linearize the
continuity equation (\ref{conti}):
\begin{equation}
\rhod_1(\rv) + \nablav\cdot\jv_1(\rv) = 0\,.
\label{contilin1}
\end{equation}
Again, we express all quantities in terms of equilibrium quantities
and the unknown quantities $\nu_1$ and $\phi_1$. According to
\Eq{defcurrent}, the current $\jv_1$ can be decomposed into
quasiparticle and superfluid contributions,
\begin{equation}
\jv_1(\rv) = \jv_{1\nu}(\rv)-\frac{\hbar}{m}
  \rho_0(\rv)\nablav\phi_1(\rv) = 0\,,
\end{equation}
where the quasiparticle contribution is given by
\begin{equation}
\jv_{1\nu}(\rv) = \int \frac{d^3p}{(2\pi\hbar)^3} \frac{\pv}{m}
  \nu_1(\rv,\pv)
\label{j1nudef}
\end{equation}
(note that only the time-odd part of $\nu_1$ contributes to the
integral, and $\nu_{1\iod} = \varrhot_{1\iod}$). The time derivative
$\nud_1$ which appears when one writes down the explicit expression
for $\rhod_1$ can be eliminated with the help of \Eq{qpvlasovlin}. As
a result, the l.h.s.\@ of the continuity equation becomes
\begin{multline}
\dot{\rho}_1+\nablav\cdot\jv_1 =\\
  \frac{A}{1+g A}\Big[\hbar\phidd_1-\dot{V}_{1\iext}
  -\Big(\frac{2\pi^2\hbar^3}{mp_F}+g\Big)
    \frac{\hbar}{m}\nablav\cdot(\rho_0\nablav\phi_1)\\
  +g\nablav\cdot\jv_{1\nu}
  +\frac{\Delta_0}{A}\nablav\cdot\int\frac{d^3p}{(2\pi\hbar)^3}
  \frac{\Delta_0}{E_0^2}\frac{\pv}{m}\nu_1\Big]\,.
\label{contilin}
\end{multline}
As noted in \Ref{UrbanSchuck}, the continuity equation is trivially
satisfied in the normal phase ($T \ge T_c$). This becomes evident if
its l.h.s.\@ is written in the form (\ref{contilin}), since in the
normal phase we have $\Delta_0 = 0$ and $A = 0$.

\subsection{Identification of important and unimportant terms}
\Eq{qpvlasovlin} is still very complicated. In order to simplify the
problem, let us look more closely at the different terms in order to
see if some of them are less important than others. The basic
assumption being that $\Delta$, $k_B T_c$, and $k_B T$ are much
smaller than $\epsilon_F$, the distribution function is sharply peaked
near the Fermi surface. Under this condition it is useful to express
the distribution function $\nu$ in terms of the variables $\rv$,
$\xi$, and $\pvh$ instead of $\rv$ and $\pv$, where
\begin{equation}
\xi = h_0(\pv,\rv) \approx v_F(\rv) [|\pv|-p_F(\rv)]\,,\quad
\pvh = \frac{\pv}{|\pv|}\,,
\end{equation}
and $v_F(\rv) = p_F(\rv)/m$. In terms of these variables, $\nu$ is
sharply peaked near $\xi = 0$, and the relevant values of $\xi$ are
of the same order of magnitude as $\Delta$, $k_B T_c$, and $k_B T$.

If $\nu_1$ is written as a function of the new variables, the Poisson
bracket on the l.h.s.\@ of \Eq{qpvlasovlin} becomes
\begin{multline}
\poisson{E_0}{\nu_1} = \frac{\Delta_0}{E_0} v_F
  \pvh\cdot(\nablav\Delta_0)\frac{\partial \nu_1}{\partial\xi}
  +\frac{\xi}{E_0}\frac{1}{p_F}(\nablav V_0)\cdot
    \frac{\partial\nu_1}{\partial\pvh}\\
  -\frac{\xi}{E_0} v_F \pvh\cdot\nablav \nu_1
  +\frac{\Delta_0}{E_0}\frac{1}{p_F}(\nablav\Delta_0)\cdot
    \frac{\partial\nu_1}{\partial\pvh}\,,
\label{poissonxi}
\end{multline}
with the short-hand notation
\begin{equation}
\frac{\partial\nu_1}{\partial\pvh} = 
  \frac{\partial\nu_1}{\partial\vartheta_p}
  \frac{\partial\pvh}{\partial\vartheta_p}
+\frac{1}{\sin^2\vartheta_p}\frac{\partial\nu_1}{\partial\varphi_p}
  \frac{\partial\pvh}{\partial\varphi_p}
\end{equation}
where $\vartheta_p$ and $\varphi_p$ denote the angles characterizing
the unit vector $\pvh = (\sin\vartheta_p\cos\varphi_p,
\sin\vartheta_p\sin\varphi_p, \cos\vartheta_p)$.

In addition to the assumption $\Delta \ll \epsilon_F$, our
semiclassical theory requires that all quantities vary slowly in
space, i.e., on a length scale $L$ which should be larger than the
coherence length $\hbar v_F/(\pi\Delta)$ \cite{FetterWalecka}. Then,
using $\Delta_0 \sim E_0 \sim \xi \sim \Delta$, $\nablav\sim 1/L$,
$\partial/\partial \xi \sim 1/\Delta$, and $\partial/\partial\pvh \sim
1$, all terms in \Eq{poissonxi} can be estimated to be of the order of
magnitude $(v_F/L) \nu_1$, except the last one, which is of the order
$(v_F/L)(\Delta/\epsilon_F) \nu_1$. Hence, the last term of
\Eq{poissonxi} is negligible.

Let us now distinguish different kinds of contributions to $\nu_1$,
depending on whether they are even or odd functions in $\xi$ and
$\pvh$:
\begin{itemize}
\item[$\nu_{1oe}$:] the part of $\nu_1$ which is odd in $\xi$ and even
in $\pvh$ describes, roughly speaking, a change of the Fermi momentum,
i.e., fluctuations of the density, and contributes to $\rho_{1\nu}$,
\begin{equation}
\rho_{1\nu} \approx \frac{mp_F}{2\pi^2\hbar^3}\int \frac{d\Omega_p}{4\pi}
  \int d\xi \frac{\xi}{E}\nu_{1oe}\,,
\end{equation}
with $d\Omega_p = \sin\vartheta_p d\vartheta_p d\varphi_p$, while its
contribution to $\Delta_{1\nu}$ is suppressed by one power of
$\Delta/\epsilon_F$ and can be neglected.
\item[$\nu_{1eo}$:] the part of $\nu_1$ which is even in $\xi$ and odd
in $\pvh$ describes a shift of the Fermi sphere and therefore
contributes to the current $\jv_{1\nu}$,
\begin{equation}
\jv_{1\nu} \approx \frac{p_F^2}{2\pi^2\hbar^3} \int \frac{d\Omega_p}{4\pi}
  \pvh\int d\xi\,\nu_{1eo}\,,
\end{equation}
and also to the other integral in the continuity equation
(\ref{contilin}).
\item[$\nu_{1ee}$:] the part of $\nu_1$ which is even in $\xi$ and in
$\pvh$ describes, roughly speaking, a local temperature fluctuation
and leads to a non-vanishing value of $\Deltat_1$ (via $\Delta_{1\nu}$),
\begin{equation}
\Deltat_1 \approx -\frac{1}{1-\varphi}\int\frac{d\Omega_p}{4\pi}\int
  d\xi\frac{\Delta_0}{E} \nu_{1ee}\,,
\end{equation}
while its contribution to $\rho_{1\nu}$ is suppressed by one power of
$\Delta/\epsilon_F$ and can be neglected.
\item[$\nu_{1oo}$:] the part of $\nu_1$ which is odd in $\xi$ and odd
in $\pvh$ gives only a negligible contribution to the current
$\jv_{1\nu}$ (suppressed by one power of $\Delta/\epsilon_F$).
\end{itemize}

If one neglects the last term in \Eq{poissonxi}, the Poisson bracket
in \Eq{qpvlasovlin} leads only to a coupling between $\nu_{1eo}$ and
$\nu_{1oe}$ and between $\nu_{1oo}$ and $\nu_{1ee}$. To be more
specific, $\nu_{1ee}$ and $\nu_{1oo}$ do not contribute to the
dynamics of $\nu_{1oe}$ and $\nu_{1eo}$. Since we are interested in
density oscillations and currents, which are determined by $\nu_{1oe}$
and $\nu_{1eo}$, we might wonder if we could disregard completely
$\nu_{1ee}$ and $\nu_{1oo}$. To that end we have to check that also on
the r.h.s.\@ of \Eq{qpvlasovlin} there is no term which couples the
undesired quantities $\nu_{1ee}$ and $\nu_{1oo}$ to $\nu_{1oe}$ or
$\nu_{1eo}$. Actually, on the r.h.s.\@ of \Eq{qpvlasovlin} there is no
term containing $\nu_{1oo}$ and only one term containing $\nu_{1ee}$,
namely the third one,
\begin{equation}
-\frac{\fp(E_0)}{m}\frac{h_0}{E_0^2}\pv\cdot\nablav
    \frac{\Delta_0\Delta_{1\nu}}{gA}
  \approx v_F\fp(E_0)\frac{\xi}{E_0^2}\pvh\cdot\nablav(\Delta_0\Deltat_1)\,.
\label{thirdterm}
\end{equation}
This term clearly contributes to $\nud_{1oo}$, but at least to leading
order in $\Delta/\epsilon_F$ it does not contribute to $\nud_{1eo}$ or
$\nud_{1oe}$. In the continuity equation (\ref{contilin}), $\nu_{1ee}$
and $\nu_{1oo}$ do not appear, i.e., the undesired quantities
$\nu_{1ee}$ and $\nu_{1oo}$ do not contribute to the dynamics of
$\phi_1$, either. We are therefore allowed to disregard them.

Now, since we are not interested any more in $\nu_{1ee}$ and
$\nu_{1oo}$, we can remove all the terms on the r.h.s.\@ of
\Eq{qpvlasovlin} which contribute only to the dynamics of these
uninteresting quantities. As mentioned above, this is the case for the
third term, \Eq{thirdterm}, which contributes only to
$\nud_{1oo}$. The last term on the r.h.s.\@ of \Eq{qpvlasovlin},
\begin{equation}
\frac{\fp(E_0)}{m}\hbar\frac{\Delta_0}{E_0}(\nablav\Delta_0)
  \cdot\nablav\phi_1
\end{equation}
can be omitted, too, since it contributes only to $\nud_{1ee}$. In
conclusion, we are left with a simplified version of \Eq{qpvlasovlin},
which reads
\begin{multline}
\nud_1-\poisson{E_0}{\nu_1} = \\
  -\frac{\fp(E_0)}{m}\Big(-\pv\cdot\nablav
    \frac{V_{1\iext}+g\rho_{1\nu}-\hbar\phid_1}{1+gA}\\
  +\frac{\Delta_0}{E_0^2}\pv\cdot\nablav
    \frac{\Delta_0(V_{1\iext}+g\rho_{1\nu}-\hbar\phid_1)}{1+gA}\\
  +\frac{\hbar}{m}\frac{h_0}{E_0}(\pv\cdot\nablav)^2\phi_1
  -\hbar\frac{h_0}{E_0}(\nablav V_0)\cdot\nablav\phi_1\Big)\,.
\label{qpvlasovsimp}
\end{multline}

\section{Test-particle method}
\label{testparticlemethod}

\subsection{Description of the method}
\label{descriptionmethod}
The aim of the present work is to solve the Vlasov-like equation
(\ref{qpvlasov}) for the quasiparticle-distribution function $\nu$
together with the continuity equation (\ref{conti}) for the phase of
the order parameter with the help of the test-particle method, in
analogy to the test-particle method which is used to solve the usual
Boltzmann equation. The basic idea of this method is to replace the
continuous distribution function $\nu(\rv,\pv)$ by a sum of delta
functions in phase space,
\begin{equation}
\nu(\rv,\pv;t) \propto \sum_i \delta[\rv-\Rv_i(t)]\delta[\pv-\Pv_i(t)]\,, 
\label{tpstandard1}
\end{equation}
corresponding to a finite number of test particles, each of which
follows the classical equation of motion
\begin{equation}
\Rvd_i = \frac{\partial E(\Rv_i,\Pv_i;t)}{\partial \Pv_i}\,,\quad
\Pvd_i = -\frac{\partial E(\Rv_i,\Pv_i;t)}{\partial \Rv_i}\,,
\label{tpstandard2}
\end{equation}
as can be seen by inserting \Eq{tpstandard1} into \Eq{qpvlasov}. Note
that, contrary to the usual test-particle method, our test particles
here cannot be identified with real particles but rather with
Bogoliubov quasiparticles. In its general form, the test-particle
method can be applied to situations far from equilibrium. However,
here we are only interested in the linear-response regime, i.e., in
the limit of small deviations from equilibrium. In this case it is
possible to formulate the method in such a way that only the classical
trajectories corresponding to the unperturbed system appear.

To that end, we make the following ansatz for the deviation of
the distribution function from equilibrium:
\begin{equation}
\nu_1(\rv,\pv;t) = -y(\rv,\pv;t) \fp[E_0(\rv,\pv)]\,.
\label{nu1ansatz}
\end{equation}
Inserting this into the linearized transport equation
(\ref{qpvlasovsimp}), we obtain the following equation of motion for
the function $y$:
\begin{equation}
\dot{y}(\rv,\pv;t)-\poisson{E_0(\rv,\pv)}{y(\rv,\pv;t)} = F(\rv,\pv;t)\,,
\end{equation}
where
\begin{multline}
F(\rv,\pv;t) = -\frac{\pv}{m}\cdot\nablav
    \frac{V_{1\iext}+g\rho_{1\nu}-\hbar\phid_1}{1+gA}\\
  +\frac{\Delta_0}{E_0^2}\frac{\pv}{m}\cdot\nablav
    \frac{\Delta_0(V_{1\iext}+g\rho_{1\nu}-\hbar\phid_1)}{1+gA}\\
  +\hbar\frac{h_0}{E_0}\Big(\frac{\pv}{m}\cdot\nablav\Big)^2\phi_1
  -\frac{\hbar}{m}\frac{h_0}{E_0}(\nablav V_0)\cdot\nablav\phi_1\,.
\label{Fexplicit}
\end{multline}
Denoting by $\Rv(\rv,\pv;t)$ and $\Pv(\rv,\pv;t)$ the
classical trajectories satisfying the equations of motion
\begin{equation}
\Rvd = \frac{\partial E_0(\Rv,\Pv)}{\partial \Pv}\,,\quad
\Pvd = -\frac{\partial E_0(\Rv,\Pv)}{\partial \Rv}
\label{trajectory}
\end{equation}
with the initial conditions
\begin{equation}
\Rv(\rv,\pv;0) = \rv\,,\quad \Pv(\rv,\pv;0) = \pv\,,
\end{equation}
one can easily show that
\begin{equation}
\frac{d}{dt}y[\Rv(\rv,\pv;t),\Pv(\rv,\pv;t);t] = 
  F[\Rv(\rv,\pv;t),\Pv(\rv,\pv;t);t]\,.
\label{eomy1}
\end{equation}

Let us now replace the quasiparticle-distribution function by $N_\nu$
delta functions in phase space. Since the order of magnitude of
$\nu_1$ is dominated by $-\fp(E_0)$, it is clear that these
delta functions should be distributed near the Fermi surface. To be
more specific, we choose $N_\nu$ points $\rv_i$, $\pv_i$ in phase
space which are distributed according to a probability density which
is proportional to $-\fp(E_0)$, in such a way that for arbitrary but
sufficiently smooth phase-space functions $\chi(\rv,\pv)$ the integral of
$\chi(\rv,\pv)$ times the function $\fp[E_0(\rv,\pv)]$ can be
approximated by
\begin{equation}
\int \frac{d^3r d^3p}{(2\pi\hbar)^3} \chi(\rv,\pv) \fp[E_0(\rv,\pv)]
  \approx -C \sum_{i=1}^{N_\nu} \chi(\rv_i,\pv_i)\,. 
\label{montecarlo}
\end{equation}
Note that, if $\rv_i$, $\pv_i$ are distributed in such a way, the same
is true for $\Rv_i(t) = \Rv(\rv_i,\pv_i;t)$, $\Pv_i(t) =
\Pv(\rv_i,\pv_i;t)$, since the quasiparticle energy $E_i =
E_0[\Rv_i(t),\Pv_i(t)]$ is a constant of the motion. In particular,
defining $y_i(t) = y[\Rv_i(t),\Pv_i(t);t]$ and using \Eq{montecarlo},
we can approximate the integral of an arbitrary function $\chi$ times the
distribution function $\nu_1$ as
\begin{equation}
\int \frac{d^3r d^3p}{(2\pi\hbar)^3} \chi(\rv,\pv) \nu_1(\rv,\pv;t)
  \approx C \sum_{i=1}^{N_\nu} y_i(t) \chi[\Rv_i(t),\Pv_i(t)]\,.
\label{nuintegral}
\end{equation}
In other words, we have replaced $\nu_1$ by
\begin{equation}
\nu_1(\rv,\pv;t)\to C \sum_{i=1}^{N_\nu} y_i(t) \delta[\rv-\Rv_i(t)]
  \delta[\pv-\Pv_i(t)]\,.
\end{equation}
According to \Eq{eomy1}, the equation of motion of the coefficients
$y_i$ is reduced to
\begin{equation}
\dot{y}_i(t) = F[\Rv_i(t),\Pv_i(t);t]\,.
\label{eomy}
\end{equation}

Above we assumed the function $\chi(\rv,\pv)$ to be sufficiently
smooth. Of course, this causes some trouble if we want to calculate
local quantities like the density or the current. For instance, we
obtain
\begin{gather}
\rho_{1\nu}(\rv) = C \sum_{i=1}^{N_\nu} y_i(t)
  \frac{\xi_i(t)}{E_i} \delta[\rv-\Rv_i(t)]\,,
  \label{rho1nuy}
\end{gather}
where $\xi_i(t) = h_0[\Rv_i(t),\Pv_i(t)]$. This result makes sense
only after the delta functions have been averaged over a volume
containing a sufficiently large number of test particles in order to
have a reasonable statistics. Supposing that this can be done, and
supposing that $V_\iext(\rv;t)$ and the phase $\phi_1(\rv;t)$ are
known, we can use the result for $\rho_{1\nu}$ in the explicit
expression for $F$ in order to obtain a system of $N_\nu$ coupled
first-order differential equations of the form (\ref{eomy}) for the
coefficients $y_i$. This represents a tremendous simplification with
respect to the original partial differential equation
(\ref{qpvlasovsimp}) in seven dimensions ($\rv$, $\pv$, and $t$).

However, the phase $\phi_1(\rv,t)$ is not known, but it has to be
determined from the continuity equation (\ref{contilin1}). This is,
again, very difficult. Hence, instead of solving the continuity
equation exactly, we make an ansatz for $\phi_1$ and determine the
parameters by minimizing the violation of the continuity equation,
\begin{equation}
\int d^3r (\rhod_1+\nablav\cdot\jv_1)^2 = \min\,,
\label{contimin}
\end{equation}
the explicit expression for $\rhod_1+\nablav\cdot\jv_1$ being given by
\Eq{contilin}. The idea is to expand $\phi_1$ on an appropriately
chosen set of orthogonal functions $\psi_n$,
\begin{equation}
\phi_1(\rv;t) = \sum_{n=1}^{N_\phi} x_n(t)\psi_n(\rv)\,.
\label{phiansatz}
\end{equation}
Inserting this ansatz into \Eq{contilin}, we see that the integral in
\Eq{contimin} depends on $x_n$ and $\ddot{x}_n$. At a given time, we
regard $x_n$ and $\dot{x}_n$ as given (e.g., at the moment when the
perturbation is switched on, we know that $x_n = \dot{x}_n =
0$). Hence, in order to have a minimal violation of the continuity
equation, we have to minimize \Eq{contimin} with respect to
$\ddot{x}_n$ by demanding
\begin{equation}
\frac{d}{d\ddot{x}_n}\int d^3r (\rhod_1+\nablav\cdot\jv_1)^2 = 0\,.
\end{equation}
At this stage it turns out to be convenient to choose the basis
functions $\psi_n$ such that they satisfy the orthogonality relation
\begin{equation}
\int d^3 r \Big(\frac{\hbar A}{1+g A}\Big)^2 \psi_n(\rv)\psi_m(\rv) =
  \delta_{nm}\,.
\label{orthogonality}
\end{equation}
Then we obtain the following differential equation for the
coefficients $x_n$:
\begin{multline}
\ddot{x}_n(t) = \sum_{m=1}^{N_\phi} a_{nm} x_m(t)
  +\int d^3r\frac{\hbar A^2\psi_n}{(1+gA)^2}\Big(\dot{V}_{1\iext}\\
    -g\nablav\cdot\jv_{1\nu}
    -\frac{\Delta_0}{A}\nablav\cdot\int \frac{d^3p}{(2\pi\hbar)^3}
      \frac{\Delta_0}{E_0^2}\frac{\pv}{m}\nu_1\Big)\,,
\label{eomxmatrix1}
\end{multline}
where $a$ is a time-independent matrix,
\begin{equation}
a_{nm} = \frac{\hbar^2}{m}\int d^3r \frac{A^2\psi_n}{(1+gA)^2}
  \Big(\frac{2\pi^2\hbar^3}{mp_F}+g\Big)
    \nablav\cdot(\rho_0\nablav\psi_m)\,.
\end{equation}
Using \Eq{nuintegral} and integrating by parts, we can rewrite
\Eq{eomxmatrix1} in a more convenient form as
\begin{equation}
\ddot{x}_n(t) = \sum_{m=1}^{N_\phi} a_{nm} x_m(t)
  +\sum_{i=1}^{N_\nu} b_{ni}(t) y_i(t)+\dot{v}_n(t)\,,
\label{eomxmatrix}
\end{equation}
where $b(t)$ denotes the matrix
\begin{equation}
b_{ni}(t) = \frac{\hbar C}{m} \Pv_i(t)\cdot\Big(\nablav
  \frac{gA^2\psi_n}{(1+gA)^2}+\frac{\Delta_0}{E_i^2}\nablav
  \frac{A\Delta_0\psi_n}{(1+gA)^2}\Big)_{\Rv_i(t)}
\end{equation}
and the vector $v$ is defined by
\begin{equation}
v_n = \hbar \int d^3r \frac{A^2 \psi_n V_{1\iext}}{(1+gA)^2}\,.
\label{vvector}
\end{equation}

Mainly for formal purposes, we note that also the equation
(\ref{eomy}) for the coefficients $y_i$ can be rewritten in matrix
notation as
\begin{multline}
\dot{y}_i(t) = \sum_{n=1}^{N_\phi} [c_{in}(t) \dot{x}_n(t) + d_{in}(t) x_n(t)]
  + f_i(t)\\
  +\sum_{j=1}^{N_\nu} g_{ij}(t) y_j(t)\,,
\label{eomymatrix}
\end{multline}
where
\begin{gather}
c_{in}(t) =
  \frac{\hbar}{m}\Pv_i(t)\cdot\Big(\nablav\frac{\psi_n}{1+gA}
    -\frac{\Delta_0}{E_i^2}\nablav
    \frac{\Delta_0 \psi_n}{1+gA}\Big)_{\Rv_i(t)}\,,
\label{cmatrix}
\\
d_{in}(t) = \frac{\hbar}{m} \frac{\xi_i(t)}{E_i}
  \Big(\frac{(\Pv_i(t)\cdot\nablav)^2\psi_n}{m}
    -(\nablav V_0)\cdot\nablav\psi_n\Big)_{\Rv_i(t)}\,,
\\
f_i(t) = -\frac{\Pv_i(t)}{m}\cdot\Big(\nablav
  \frac{V_{1\iext}}{1+gA}-\frac{\Delta_0}{E_i^2}\nablav
    \frac{\Delta_0 V_{1\iext}}{1+gA}\Big)_{\Rv_i(t)}\,,
\label{fvector}
\end{gather}
and
\begin{multline}
g_{ij}(t) = -g C \frac{\xi_j(t)}{E_j}\frac{\Pv_i(t)}{m}\cdot\Big(\nablav
  \frac{\deltat[\rv-\Rv_j(t)]}{1+gA}\\
  -\frac{\Delta_0}{E_i^2}\nablav
    \frac{\Delta_0 \deltat[\rv-\Rv_j(t)]}{1+gA}\Big)_{\Rv_i(t)}\,.
\label{gijsmeared}
\end{multline}
In the latter equation, $\deltat$ denotes a kind of ``smeared'' delta
function which accounts for the averaging mentioned below
\Eq{rho1nuy}.

However, as mentioned above, \Eqs{eomymatrix} and (\ref{gijsmeared})
will be used for formal purposes only. In practice, it is much faster
to calculate $\rho_{1\nu}(\rv)$ after each time step on a discrete
mesh, and to interpolate the stored values when performing the next
time step for the coefficients $y_i$. For the calculation of
$\rho_{1\nu}(\rv)$, we replace the delta function in \Eq{rho1nuy} by a
Gaussian having a width $d_\rho$.

In summary, the coupled system of partial differential equations,
namely the transport equation for the distribution function $\nu_1$
and the continuity equation for the phase $\phi_1$ [\Eqs{qpvlasovsimp}
and (\ref{contilin1})], has been replaced by a coupled system of
ordinary linear differential equations for the coefficients $y_i$ and
$x_n$ [\Eqs{eomymatrix} and (\ref{eomxmatrix})], which can formally
be written as
\begin{equation}
\frac{d}{dt}\Bigg(\begin{matrix} x(t)\\ \dot{x}(t)\\ y(t)
  \end{matrix}\Bigg) 
= \Bigg(\begin{matrix} 0& 1& 0\\ a& 0& b(t)\\ d(t)& c(t)&
  g(t) \end{matrix}\Bigg)
  \Bigg(\begin{matrix}x(t)\\ \dot{x}(t)\\ y(t)\end{matrix}\Bigg)
  +\Bigg(\begin{matrix}0\\ \dot{v}(t)\\ f(t)\end{matrix}\Bigg)\,.
\label{eommatrix}
\end{equation}

\subsection{Trajectories of the test particles}
In practice, the solution of the classical equations of motion for the
test particles, \Eqs{trajectory}, faces us with some unusual features
which are not present with the usual Newtonian equations of
motion. Note that we are not dealing with ordinary particles but with
Bogoliubov quasiparticles, which have some surprising properties. For
instance, $E_i$ being a constant of the motion and $E_i^2 =
\xi_i^2+\Delta_0^2(\rv_i)$, it is evident that the energy $\xi_i$
cannot be conserved if the gap $\Delta_0$ depends on $\rv$. In
particular, when a test particle with quasiparticle energy $E_i$
reaches the surface where $\Delta_0(\rv) = E_i$, it is reflected
(Andreev reflection). During this reflection, the momentum $\Pv_i$
stays almost constant, but the energy $\xi_i$ changes its sign (i.e.,
a particle is transformed into a hole or vice versa), such that the
velocity $\vv_i = \partial E_i/\partial\Pv_i = (\xi_i/E_i) \Pv_i/m$ is
reversed. As a consequence, the quasiparticle is reflected into the
direction where it came from, which is very surprising if the incident
angle is different from $90^\circ$.

In order to find the test-particle trajectories numerically, it does
not seem very efficient to start directly from \Eqs{trajectory}, since
a small numerical error in the momentum of the order of $\delta P/P
\sim \Delta/\epsilon_F$ would immediately lead to a completely wrong
behavior. It is therefore advantageous to make use of the variable
$\xi_i$, whose equation of motion reads
\begin{equation}
\dot{\xi_i} = -\frac{\Delta_0(\Rv_i)}{E_i} \frac{\Pv_i}{m} \cdot
  \nablav \Delta_0(\Rv_i)\,.
\end{equation}
Solving this equation together with the equations for $\Rv_i$ and
$\Pv_i$, we can correct $\Pv_i$ after each time step according to
\begin{equation}
\Pv_i^{\mathit{corr.}} = \frac{\Pv_i}{|\Pv_i|}\sqrt{2m\xi_i+p_F^2(\Rv_i)}\,.
\end{equation}

In practice, the variable $\xi_i$ also allows us to introduce a very
reliable method for determining the step size. Let us denote by
$\xi_i^\prime$ the result we obtain after one time step of size
$\delta t$, and by $\xi_i^{\prime\prime}$ the result we obtain after
two time steps of size $\delta t/2$ each. Then the quantity $\delta
t|\xi^\prime-\xi^{\prime\prime}|$ is a measure for the numerical error
and can be used for adapting the step size $\delta t$ to the
situation. It turns out that the step size has to become very small
only during Andreev reflection.

Now let us give some examples for typical test-particle
trajectories. For that purpose, let us restrict ourselves to the most
simple case which is a spherical harmonic trap,
\begin{equation}
V_{0\iext}(\rv) = \frac{1}{2}m\Omega^2 r^2\,.
\label{sphericaltrap}
\end{equation}
This potential defines the so-called trap units, i.e., energies are
measured in units of $\hbar\Omega$, temperatures in units of
$\hbar\Omega/k_B$, lengths in units of $l_{ho} =
\sqrt{\hbar/(m\Omega)}$, etc. In this example, due to spherical
symmetry, not only the quasiparticle energy $E$, but also the angular
momentum $\Lv = \rv\times \pv$ of a test particle is a constant of the
motion.

Within the local-density approximation (LDA)
\cite{Houbiers,GrassoUrban}, the density $\rho_0(r)$ has its maximum
at the center of the trap and vanishes approximately (except for very
small temperature effects) at the Thomas-Fermi radius $R_{TF} =
\sqrt{2\mu/(m\Omega^2)}$. The gap $\Delta_0(r)$ has its maximum at the
center of the trap, too, and goes to zero at some critical radius
$R_c$ which is temperature dependent and determined by the equation $T
= T_c(R_c)$. In order to avoid numerical problems arising from the
infinite derivative of $\Delta_0(r)$ at $r = R_c$, we convolute the
LDA result for $\Delta_0(r)$ with a Gaussian of width $d_\Delta$. In
fact, this is more realistic than the LDA result since the exact
solution of the BdG equations also leads to a gap $\Delta_0(r)$ which
has an exponential tail
\cite{GrassoUrban,BaranovPetrovGL,BruunCastin}. As parameters we
choose $\mu = 32\,\hbar\Omega$, $g = -\hbar^2 l_{ho}/m$, and $T =
1.4\, \hbar\Omega/k_B$. The corresponding number of atoms in the trap
is approximately $1.7\times 10^4$. For these parameters quantum
mechanical (BdG, QRPA) results are available for comparison. The width
$d_\Delta$ is chosen such as to optimize the agreement with the BdG
gap, which for the present parameters is achieved with $d_\Delta =
l_{ho}$.

\begin{figure}
\includegraphics[width=7cm]{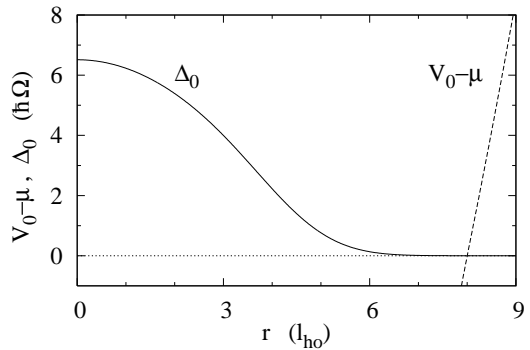}
\caption{Gap $\Delta_0(r)$ (solid line) and potential $V_0(r)-\mu$
(dashed line) for the case of a spherical trap with frequency
$\Omega$, chemical potential $\mu = 32\,\hbar\Omega$, coupling
constant $g = -\hbar^2 l_{ho}/m$, and temperature $k_B T =
1.4\,\hbar\Omega$. $\Delta_0$ and $V_0-\mu$ are in units of
$\hbar\Omega$, $r$ is in units of the oscillator length
$l_{ho}$. Roughly speaking, these two curves determine the classically
allowed region for a quasiparticle with given energy
$E$.}\label{figgroundstate1}
\end{figure}

In \Fig{figgroundstate1} we show the corresponding gap $\Delta_0(r)$
as a function of the distance $r$ from the center of the trap. From
this figure it is evident that due to the condition $E\geq
\Delta_0(r)$, the relevant quasiparticles (having $E\lesssim k_B T =
1.4\,\hbar\Omega$) are excluded from the region $r\lesssim
4 l_{ho}$. In addition to the gap, we display the potential
$V_0(r)-\mu$, since the motion of a quasiparticle with given energy
$E$ and angular momentum $\Lv$ is also limited by the condition
$\sqrt{E^2-\Delta_0^2(r)}-\Lv^2/2mr^2\geq V_0(r)-\mu$. It has been
shown that also within the fully quantum-mechanical BdG theory the
lowest-lying quasiparticle states are localized in this region
\cite{BruunHeiselberg}. In our example, the motion of the relevant
quasiparticles is restricted to the region $4\lesssim
r/l_{ho}\lesssim 8$.

\begin{figure}
\includegraphics[height=4.7cm]{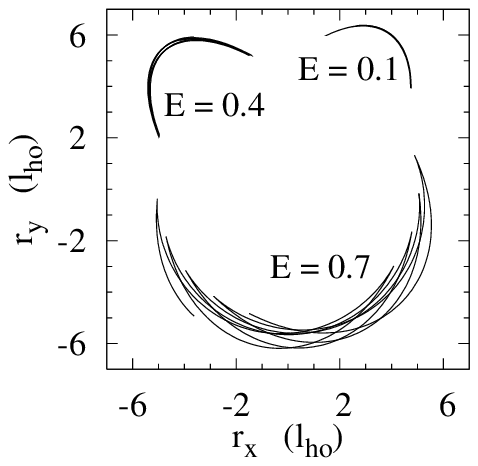}
\includegraphics[height=4.7cm]{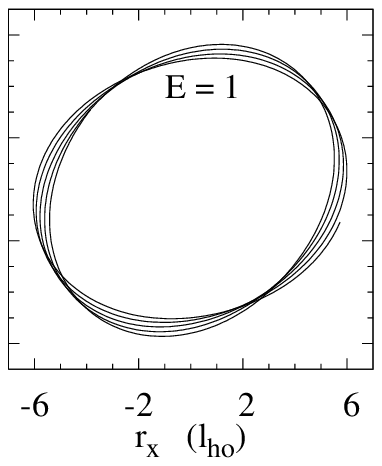}
\caption{Four examples of quasiparticle trajectories in a trap with
parameters given below \Fig{figgroundstate1}. The three trajectories
shown in the left panel belong to quasiparticles with $E =
0.1\,\hbar\Omega$, $0.4\,\hbar\Omega$, and $0.7\,\hbar\Omega$,
respectively. The trajectory displayed in the right panel belongs to a
quasiparticle with $E = \hbar\Omega$.}\label{figtraj}
\end{figure}

Most of these quasiparticles will undergo Andreev reflection. Their
trajectories are approximately described by an ellipse which is cut at
the points where $\Delta_0(r) = E$. If $E\ll \epsilon_F$, the
quasiparticle will move hence and forth on the same partial
ellipse. Such trajectories with $E = 0.1\,\hbar\Omega$ and
$0.4\,\hbar\Omega$ are shown in the left panel of
\Fig{figtraj}. However, if the quasiparticle energy is higher, the
change in energy from $\xi\approx E$ to $\xi\approx -E$ (or vice
versa) during the Andreev reflection results in a change of momentum
which is no more negligible. Then, due to angular momentum
conservation, the angle of reflection is slightly different from the
angle of incidence, and the whole trajectory is precessing. An example
for such a trajectory with $E = 0.7\,\hbar\Omega$ is also shown in the
left panel of \Fig{figtraj}. A completely different picture arises if
the initial conditions are such that the quasiparticle does never
reach the point where $\Delta(r) = E$. Then the trajectory is just a
precessing, slightly deformed ellipse, as shown in the right panel of
\Fig{figtraj} for the case of a trajectory with $E =
\hbar\Omega$. There is a striking analogy between these trajectories
and the ``glancing'' orbits discussed, e.g., in \Ref{BruderImry} in
the context of a superconducting cylinder which is coated by a
normal-metal layer.
\subsection{Distribution of test particles in phase space}
In \Sec{descriptionmethod} we supposed that one can generate a
distribution of points $\rv_i$, $\pv_i$ in phase space such that
\Eq{montecarlo} is approximately satisfied for sufficiently smooth
functions $\chi(\rv,\pv)$. In practice, this distribution is obtained
in two steps. First we generate the coordinates $\rv_i$, and in a
second step the momenta $\pv_i$.

The mean density of test particles at a certain point $\rv$ is given
by
\begin{equation}
n(\rv) = \sum_{i=1}^{N_\nu} \deltat(\rv_i-\rv)\,,
\end{equation}
where $\deltat$ denotes a smeared delta function in order to account
for the averaging. Using \Eq{montecarlo}, we conclude
\begin{equation}
n(\rv) = -\frac{1}{C} \int \frac{d^3p}{(2\pi\hbar)^3}
  \fp[E_0(\rv,\pv)]\equiv \frac{w(\rv)}{C}\,.
\label{ntestparticle}
\end{equation}
The algorithm for the generation of the coordinates $\rv_i$ is now
very simple. First we look for the maximum $w_\imax$ of the function
$w(\rv)$. Defining $P(\rv) = w(\rv)/w_{\imax}$, we obtain a function
whose values lie between 0 and 1. Then we generate uniformly
distributed random points $\rv_k$ in a volume which contains the whole
system, and retain each point with the probability $P(\rv_k)$, until
the desired number of points, $N_\nu$, is reached.

The formula (\ref{ntestparticle}) for the test-particle density $n(\rv)$
can also be used for the determination of the normalization constant
$C$. Integrating $n(\rv)$ over space, we must recover the total number of
test particles. This implies
\begin{equation}
C = \frac{1}{N_\nu} \int d^3r\, w(\rv)\,.
\end{equation}

Now we turn to the distribution of the momenta $\pv_i$. It is evident
that the angular distribution of the momenta is isotropic, i.e., the
interesting part of the problem is the distribution of the absolute
values, $p_i = |\pv_i|$, which is, of course, directly related to the
distribution of the energies $\xi_i$. Let us define the mean number of
test particles per energy and volume
\begin{equation}
n(\rv,\xi) = \sum_{i=1}^{N_\nu} \deltat(\rv_i-\rv)
  \deltat(\xi_i-\xi)\,.
\end{equation}
Again, with the help of \Eq{montecarlo}, this becomes
\begin{equation}
n(\rv,\xi) = -\frac{1}{C} \frac{m p_\xi}{2\pi^2\hbar^3}
  \fp(E_\xi)\,,
\end{equation}
with $p_\xi = \sqrt{2m\xi+p_F^2(\rv)}$, i.e., for given spatial
coordinates $\rv$, the probability density for finding a particle at
energy $\xi$ is proportional to $-p_\xi\fp(E_\xi)$. Such a
distribution can be generated in the following way. Starting from
random numbers $z_k$ which are uniformly distributed in the interval
$(0,1)$, it is straight-forward to show that the energies
\begin{equation}
\xi_k = T \ln \frac{z_k}{1-z_k}
\label{xidistrib}
\end{equation}
are distributed according to the probability density $-\fp(\xi)$. It
is evident that negative energies with $\xi < -\epsilon_F(\rv)$ have
to be removed. Furthermore, it is preferable to cut the distribution
at energies which lie too far away from the Fermi surface, e.g.,
$|\xi| > 15 T$ (the probability that this happens is less than
$10^{-6}$). The momenta $p_\xi$ are thus limited by $p_\imax =
\sqrt{30mT+p_F^2(\rv)}$, and the function defined by $P(\xi) =
p_\xi\fp(E_{\xi})/p_\imax\fp(\xi)$ cannot become greater than 1 and
can serve as a probability. If we retain each energy $\xi_k$ generated
according to \Eq{xidistrib} with the probability $P(\xi_k)$, the
remaining energies are distributed according to the desired
distribution.

\begin{figure}
\includegraphics[width=7cm]{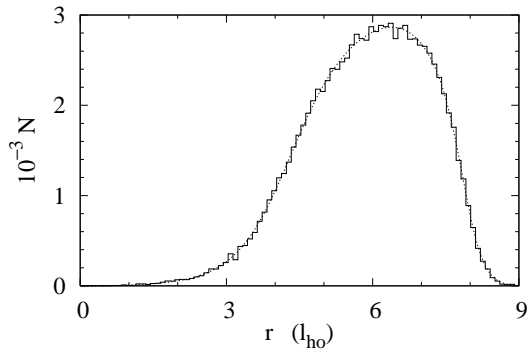}
\caption{Radial distribution of $10^5$ test particles in a trap with
parameters given below \Fig{figgroundstate1}, counted in bins of
size $\delta r = 0.1\,l_{ho}$. For comparison, the dotted curve
represents the ideal distribution according to
\Eq{ntestparticle}.}\label{figtestparticles}
\end{figure}

In order to give an illustration for the resulting distribution of
test particles, we show in \Fig{figtestparticles} the radial
distribution of $N_\nu = 10^5$ test particles in a trap with the same
parameters as in \Fig{figgroundstate1}. In agreement with what we
discussed in the preceding subsection, we see that the test particles
are mainly located in the region $4\lesssim r/l_{ho}\lesssim 8$,
corresponding to the region where the system is mainly normal
fluid. Due to the angular average the statistical fluctuations around
the ideal distribution, \Eq{ntestparticle}, which is represented by
the dotted line, are very small. We verified that, apart from the
statistical fluctuations, our test-particle distribution stays
constant, which is a good numerical test of both the initial
test-particle distribution and of the test-particle trajectories.
\subsection{Initial condition}
In the linear response regime, as the name implies, the response to a
time-dependent perturbation of the form $V_1(\rv;t) = \Vhat_1(\rv)
f(t)$, with an arbitrary time dependence $f(t)$, can be obtained as
convolution of $f(t)$ with the response to a delta function in
time. It is therefore sufficient to study perturbations of the form
\begin{equation}
V_{1\iext}(\rv;t) = \Vhat_1(\rv) \delta(t)\,.
\label{perturbationdelta}
\end{equation}
We thus set the inhomogeneous terms in \Eq{eommatrix} to $\dot{v}(t) =
\vhat \dot{\delta}(t)$ and $\dot{f}(t) = \fhat \delta(t)$,
respectively, $\vhat$ and $\fhat$ being defined analogously to
\Eqs{vvector} and (\ref{fvector}) but with $V_{1\iext}$ replaced by
$\Vhat_1$.

Assuming that the system was in equilibrium before this perturbation,
we may ask the question: What are the values of the coefficients $y_i$
and $x_n$ immediately after the perturbation, i.e., at infinitesimally
small $t > 0$? This question can be answered exactly, since during the
infinitesimal period where the perturbation is active, the matrix in
\Eq{eommatrix} can be regarded as time-independent. Integrating
\Eq{eommatrix} over time from $-t_0$ to $t_0$, we obtain in the limit
$t_0\to 0$ 
\begin{equation}
\lim_{t_0\to 0} \Bigg(\begin{matrix}x(t_0)\\ \dot{x}(t_0)\\ y(t_0)
  \end{matrix}\Bigg)
= \Bigg(\begin{matrix}\vhat\\ 0\\ c\vhat+\fhat
  \end{matrix}\Bigg)\,.
\label{initial}
\end{equation}

Let us now assume that the function $\Vhat_1$ lies in the space
spanned by the functions $\psi_n$. Then it is evident that the
corresponding linear combination is given by the coefficients $\vhat_n$,
i.e.,
\begin{equation}
\Vhat_1(\rv) = \hbar \sum_{n=1}^{N_\phi} \vhat_n \psi_n(\rv)\,.
\label{spanv1}
\end{equation}
Note that the functions $\psi_n$ do not necessarily have to have this
property. For example, we could define a basis of functions satisfying
the orthogonality relation (\ref{orthogonality}) and maybe even a
suitably defined completeness relation if $N_\phi\to\infty$, but which
all vanish identically outside the superfluid region, i.e., in the
region where $\Delta_0 = 0$ (and $A = 0$). \Eq{spanv1} would then be
satisfied inside the superfluid region, but not outside. Hence, it is
an additional requirement for the choice of the functions $\psi_n$.
Combining \Eqs{initial} and (\ref{spanv1}), we find
\begin{equation}
\lim_{t_0\to 0} \phi_1(\rv;t_0) = \frac{1}{\hbar}\Vhat_1(\rv)\,.
\label{initialphi}
\end{equation}

\Eq{spanv1} also leads to a simplification of the initial value of the
coefficients $y_i$ and the quasiparticle-distribution function. Using
the explicit expressions for the matrix $c$ and the vector $\fhat$
[\Eqs{cmatrix} and (\ref{fvector}) with $V_{1\iext}$ replaced by
$\Vhat_1$], we obtain from the third line of \Eq{initial}
\begin{multline}
\lim_{t_0\to 0} y_i(t_0) = \sum_{n=1}^{N_\phi} c_{in} \vhat_n + \fhat_i\\
  = \frac{\pv_i}{m}\cdot\Big(\nablav
  \frac{\hbar\sum_{n=1}^{N_\phi} \vhat_n \psi_n-\Vhat_1}{1+gA}\\
    -\frac{\Delta_0}{E_i^2}\nablav
      \frac{\Delta_0 (\hbar\sum_{n=1}^{N_\phi}\vhat_n\psi_n-\Vhat_1)}{1+gA}
      \Big)_{\rv_i}\,.
\label{initialy}
\end{multline}
As a consequence, if \Eq{spanv1} is satisfied, the initial values of
the coefficients $y_i$ vanish, which implies
\begin{equation}
\lim_{t_0\to 0} \nu_1(\rv,\pv;t_0) = 0\,.
\label{initialnu}
\end{equation}

In fact, the simple result of this subsection, which is summarized in
\Eqs{initialphi} and (\ref{initialnu}), could have been anticipated
without any calculation. The effect of a perturbation of the form
(\ref{perturbationdelta}) is to give a particle at position $\rv$ a
kick
\begin{equation}
\delta\pv = -\int dt \nablav V_{1\iext}(\rv;t) = -\nablav \Vhat_1(\rv)\,.
\end{equation}
Since this kick does not depend on the momentum of the particle, the
local Fermi sphere is shifted as a whole, there is no change in
density and no Fermi surface deformation. Within the present
theoretical framework, Cooper pairs are not broken either, they just
acquire a center of mass momentum. Thus, the distribution function in
the local rest frame stays unchanged ($\nu_1 = 0$), and the collective
velocity is given by $\vv_\icoll = -(\hbar/m)\nablav\phi_1 =
-(1/m) \nablav \Vhat_1$.

Note, however, that in reality a perturbation which has the form of a
short pulse would lead to much more complicated effects (e.g., pair
breaking). Since our semiclassical description requires that the time
dependence of the perturbation is slow, our formal result for a
delta-like excitation becomes physically meaningful only after it has
been convoluted with a function $f(t)$ which varies slowly in time. In
other words, we can only calculate the low-frequency part of the
response function.
\section{First results}
\label{firstresults}
In this section we will discuss first numerical results which have
been obtained using the test-particle method. Our intention here is to
see whether this method is in principle capable to describe the most
important features of collective excitations in superfluid trapped
Fermi gases. To that end, we will study the quadrupole excitation of a
spherical system, which is excited by
\begin{equation}
\Vhat_1(\rv) = \alpha\,m\Omega (2r_z^2-r_x^2-r_y^2)
\label{Vperturbation}
\end{equation}
(the factor $m\Omega$ has been introduced in order to make the
coefficient $\alpha$ dimensionless).

For practical purposes, we will make an additional approximation: We
will restrict our ansatz for the phase, \Eq{phiansatz}, to only one or
two functions $\psi_n$. It is clear from rotational symmetry that in
the case of a quadrupole excitation of the form (\ref{Vperturbation})
the most general form the phase can have is
\begin{equation}
\phi_1(\rv) = \Phi(r) [2r_z^2-r_x^2-r_y^2]\,,
\end{equation}
such that that the functions $\psi_n$ can be written as
\begin{equation}
\psi_n(\rv) = \Psi_n(r) [2r_z^2-r_x^2-r_y^2]\,.
\end{equation}
It is known from superfluid hydrodynamics that at zero temperature the
velocity field is essentially linear in the coordinates, i.e., the
function $\Phi(r)$ is almost constant. As a first guess we will assume
that this is still true at non-zero temperature, and hence we will
take only one single function ($N_\phi=1$) in the ansatz
(\ref{phiansatz}) for the phase, $\Psi_1 = \mathit{const}$. The
proportionality constant will be determined from the normalization
condition (\ref{orthogonality}).

Such a restricted ansatz means of course that the continuity equation
will not be exactly satisfied in the superfluid region (remember that
outside the superfluid region the phase has no effect whatsoever). We
will therefore improve this initial ansatz by including a second
function ($N_\phi = 2$) which allows to modulate $\Phi(r)$ in the
superfluid region.

The first idea one might have is to use for $\Psi_n(r)$ polynomials in
$r^2$ and to orthogonalize the resulting functions $\psi_n$. However,
it turns out that this leads to numerical instabilities due to the
fast growing of the resulting polynomials outside the superfluid
region. Let us explain this effect in some more detail. As seen from
the transport equation for the quasiparticle-distribution function,
the phase $\phi_1$ outside the superfluid region enters directly the
dynamics of $\nu_1$. Although the net effect of the phase and of the
quasiparticle motion should be independent of the choice of $\phi_1$
outside the superfluid region, each of these contributions depends on
this choice. If $\phi_1$ changes too rapidly, the numerical solution
of the equation of motion for the coefficients $y_i$ becomes less
accurate and the cancellation of the two effects does not work any
more.

We therefore have to look for functions $\Psi_n$ which are linearly
independent inside the superfluid region, but which do not grow
outside. Here we will choose the functions $\tilde{\Psi}_1(r) = 1$ and
$\tilde{\Psi}_2(r) = [1-\varphi(r)]^2$, where $\varphi(r)$ is the
function defined in \Eq{defphi}. The function $\tilde{\Psi}_2(r)$ has
its maximum in the center of the trap and goes smoothly to zero at the
boundary of the superfluid region. From $\tilde{\Psi}_1$ and
$\tilde{\Psi}_2$ the functions $\Psi_1$ and $\Psi_2$ are determined
according to the orthogonality condition (\ref{orthogonality}) with
the help of the Gram-Schmidt orthogonalization method. As we will see,
the results obtained with $N_\phi = 1$ and $N_\phi = 2$ are very
similar and we therefore claim that they would not change a lot if we
included additional functions.

\begin{figure}
\includegraphics[width=7cm]{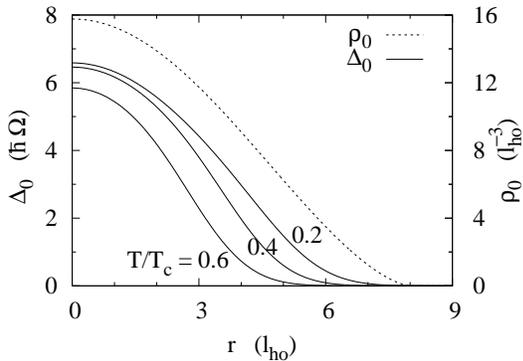}
\caption{Density profile $\rho_0(r)$ (dashed line) and gap
  $\Delta_0(r)$ (solid lines) in a spherical harmonic trap containing
  $1.7\times10^4$ atoms ($\mu = 32\,\hbar\Omega$, interaction strength
  $g = -\hbar^2 l_{ho}/m$). The gap is displayed for three different
  temperatures, $T/T_c = 0.2$, $0.4$, and $0.6$, while the density
  profile is practically independent of $T$.}
  \label{figgroundstate2}
\end{figure}

Let us now present the results. As in the examples shown in the
preceding section, we consider a spherical harmonic trap with $\mu =
32\hbar\Omega$, containing approximately $1.7\times 10^4$ atoms. The
resulting density profile $\rho_0(r)$ is shown in
\Fig{figgroundstate2} as the dashed line. The critical temperature
within LDA is $T_c = T_c(r=0) \approx 3.9\, \hbar\Omega/k_B$. As before, the
LDA result for the gap $\Delta_0(r)$ is convoluted with a Gaussian
having a width $d_\Delta = l_{ho}$. We will study the quadrupole mode
for three different temperatures, $T/T_c = 0.2$, $0.4$, and $0.6$. The
equilibrium gap $\Delta_0(r)$ for these three temperatures is also
displayed in \Fig{figgroundstate2}.

\begin{figure}
\includegraphics[width=7cm]{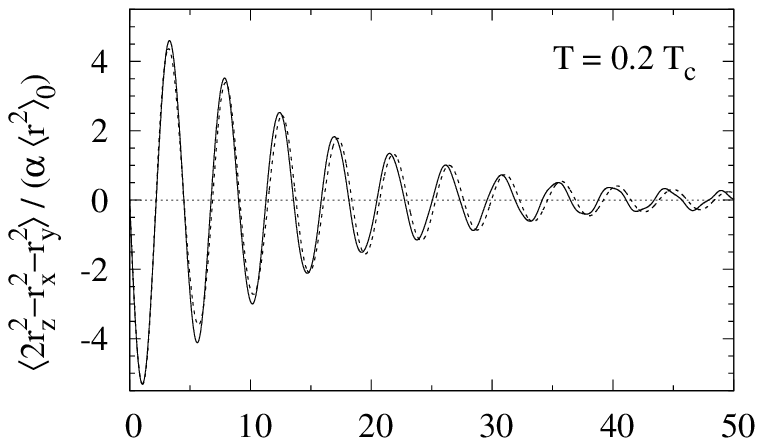}\\
\includegraphics[width=7cm]{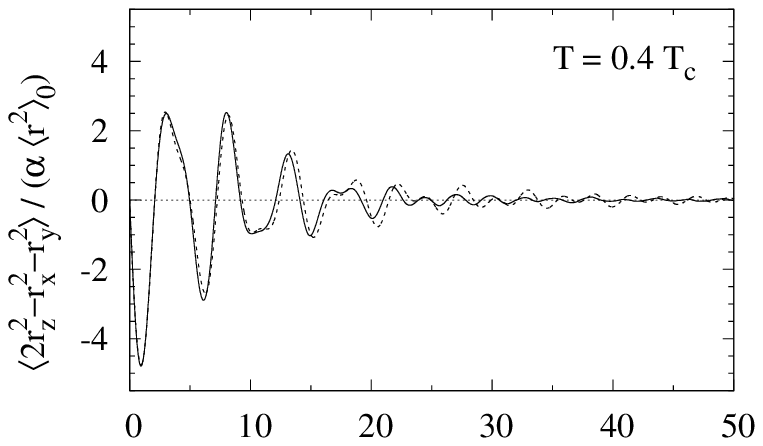}\\
\includegraphics[width=7cm]{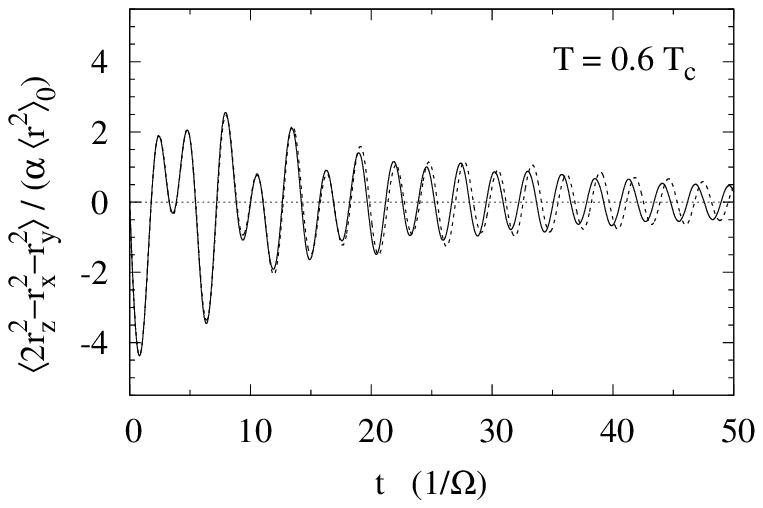}\\
\caption{Time dependence of the quadrupole deformation after a
  delta-like perturbation at $t = 0$. The parameters are the same as in
  \Fig{figgroundstate2}, the three panels correspond, from top to
  bottom, to $T/T_c = 0.2$, $0.4$, and $0.6$. The dashed and solid lines
  correspond to $N_\phi = 1$ and $N_\phi = 2$, respectively.}
  \label{figq}
\end{figure}

\begin{figure}
\includegraphics[width=7cm]{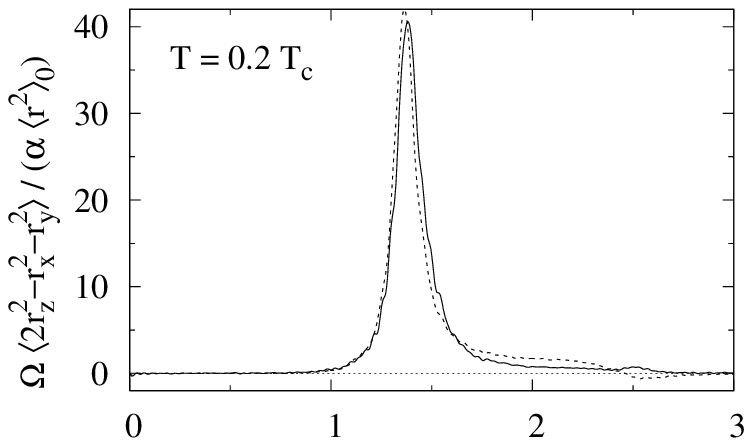}\\
\includegraphics[width=7cm]{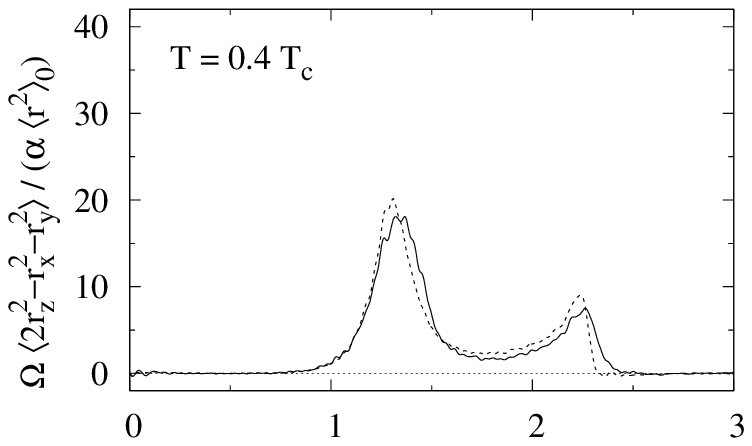}\\
\includegraphics[width=7cm]{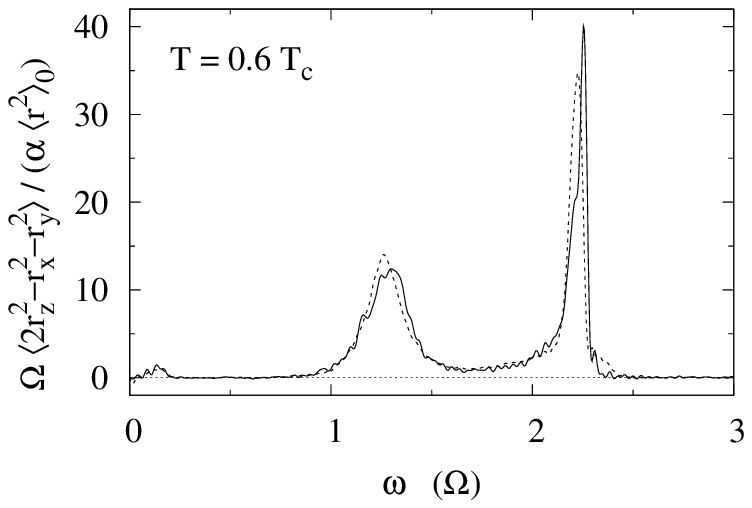}\\
\caption{Fourier transforms of the quadrupole responses shown in
  \Fig{figq}.}\label{figqfft}
\end{figure}

After the system is excited, its shape will oscillate. A measure for
this quadrupole deformation is the ratio
\begin{equation}
\frac{\langle 2r_z^2-r_x^2-r_y^2\rangle}{\langle r^2\rangle_0}\,,
\end{equation}
where $\langle r^2\rangle_0$ denotes the mean square radius in
equilibrium, which in the present case has the value $\langle
r^2\rangle_0 \approx 23\, l_{ho}^2$. In the linear response, the
quadrupole deformation is of course proportional to the strength of
the perturbation, and we therefore divide our results by this strength
[denoted $\alpha$ in \Eq{Vperturbation}]. In our simulation we use
$N_\nu = 10^5$ test particles, the width of the Gaussians used for
smearing $\rho_{1\nu}$ (see \Sec{descriptionmethod}) is set to $d_\rho
= l_{ho}$. In \Fig{figq} we display the time dependence of the
quadrupole deformation after the perturbation for the three
temperatures mentioned before. The corresponding spectra, obtained by
Fourier transformation, are shown in \Fig{figqfft}. The results for
the two cases $N_\phi = 1$ and $N_\phi = 2$ are displayed as dashed
and solid curves, respectively. In all cases the two curves are in
reasonable agreement, such that we can say that the use of $N_\phi=2$
independent functions in the ansatz for the phase is sufficient.

We see that the temperature dependence of the spectrum is highly
non-trivial. At low temperatures (upper panel of \Fig{figqfft}), we
see essentially the hydrodynamic quadrupole mode, which at zero
temperature lies at $\omega = \sqrt{2} \Omega$
\cite{BruunClark,BaranovPetrovCM,StringariS} and which is now damped
as a consequence of its coupling to the normal component. At higher
temperatures (middle of \Fig{figqfft}), a second peak builds up in the
spectrum, corresponding to the quadrupole mode in the normal phase,
which lies slightly above $\omega = 2 \Omega$ \cite{StringariN} (for
the present set of parameters, its frequency is $\omega \approx
2.2\,\Omega$ \cite{GrassoKhan}). As the temperature approaches $T_c$
(lower panel of \Fig{figqfft}), the strength contained in this second
peak increases, while the hydrodynamic mode, whose frequency is
slightly shifted downwards, disappears. These findings are in good
agreement with quantum mechanical QRPA calculations \cite{GrassoKhan}.

We note that the damping width of the hydrodynamic mode at low
temperature is now comparable with that found within the QRPA and much
stronger than that found in our previous work \cite{UrbanSchuck},
where we replaced the gap $\Delta_0(r)$ by a constant. The reason is
in fact very simple: With a constant gap, the fraction $\rho_n/\rho_0$
of the normal component is independent of $r$, whereas in the case of
an $r$-dependent gap the normal component in the outer part of the
system is already important at very low temperatures \cite{Urban}.

As the temperature $T$ approaches $T_c$, the quadrupole mode of the
normal phase (that at $\omega = 2.2\,\Omega$) becomes undamped, as it is
the case within the QRPA. However, even though collisions are strongly
suppressed at these low temperatures, it should be kept in mind that
the collision term, which is neglected in the present work, is
non-zero and its inclusion would lead to a finite lifetime of this
oscillation, too.

\begin{figure}
\includegraphics[width=7cm]{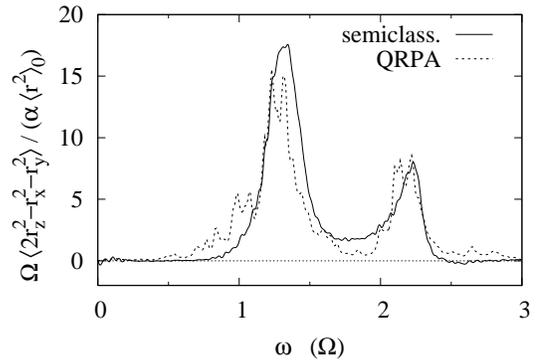}
\caption{Quantitative comparison between the semiclassical and the
QRPA result for the quadrupole excitation spectrum of a system with
$\mu = 32\hbar\Omega$, $g = -0.965\, \hbar^2 l_{ho}/m$, $k_B T =
1.4\,\hbar\Omega$. The semiclassical result (solid line) was obtained
with $N_\phi = 2$ basis functions for the phase. The QRPA result
(dotted line) was taken from \Ref{GrassoKhan}.}\label{figqrpa}
\end{figure}

Finally, let us compare our semiclassical results more quantitatively
to quantum mechanical QRPA results. In \Fig{figqrpa} we show the QRPA
result of \Ref{GrassoKhan} for the quadrupole excitation spectrum
(dotted line) together with the semiclassical result we obtain with
the same parameters (solid line). As one can see, the total
normalization and the relative weights of the two peaks are in
reasonable agreement. Also the widths of the QRPA peaks are well
reproduced by the semiclassical calculation. The main differences are
that within the semiclassical calculation the two peaks lie a bit too
high and that they are not as well separated as within the QRPA. A
comparison of the two curves for $N_\phi = 1$ and $N_\phi = 2$ in the
$T/T_c = 0.4$ case shown in the middle of \Fig{figqfft}, whose
parameters are quite close to those of \Fig{figqrpa}, suggests that
the latter effect might be partly due to the restricted ansatz for the
phase. However, as one can deduce from the irregular structure of the
QRPA spectrum, even in a system with 32000 atoms shell effects, i.e.,
effects which depend on the discrete single-particle spectrum, are
still quite pronounced. It is clear that such effects cannot be
reproduced within a semiclassical calculation. In this sense the
agreement between the two spectra is very satisfactory, in particular
since one can assume that the shell effects decrease with increasing
number of particles.

\section{Conclusions}
\label{conclusions}
In this paper, we developed a numerical test-particle method for
solving the semiclassical transport equations for an ultracold trapped
Fermi gas in the BCS phase in the collisionless limit. These transport
equations take into account the coupling between the dynamics of the
Cooper pairs (superfluid component) and the thermally excited
Bogoliubov quasiparticles (normal component). We developed the method
for the case of small deviations from equilibrium, so that the
test-particle trajectories can be calculated in the equilibrium
state. Since the test-particles describe Bogoliubov quasiparticles
rather than real particles, the trajectories have very unusual
properties compared with the trajectories one has to deal with when
applying the test-particle method to the normal Vlasov equation. Our
test particles can have the character of particles as well as holes,
depending on whether their energy $\xi$ is positive or negative, and
they can also be transformed from the one into the other if they hit
the region where the gap $\Delta$ becomes larger than their
quasiparticle energy $E$ (Andreev reflection). Another complication
as compared with the normal Vlasov equation is that the dynamics of
the quasiparticles is coupled to the collective motion of the
superfluid component, which is described by the phase $\phi$ of the
order parameter. This phase has to be determined simultaneously with
the evolution of the quasiparticle-distribution function by solving
the continuity equation. In the present work, we make an ansatz for
$\phi$ with time-dependent coefficients, leading to an approximate
solution of the continuity equation.

As a first application, we calculated the response of a gas trapped in
a spherical trap to a delta-like perturbation of quadrupole
form. After this perturbation, the shape of the gas shows a damped
oscillation. At low temperatures, this oscillation is just the
hydrodynamic quadrupole mode which is damped by its coupling to the
normal component. With increasing temperatures, the extension of the
normal component increases, and, as a consequence, the normal
component can perform its own quadrupole oscillation. Since the
frequency of the quadrupole mode in the normal collisionless Fermi gas
is higher than that of the hydrodynamic mode, this leads to a two-peak
structure in the response function. As the temperature approaches
$T_c$, the strength of the hydrodynamic mode disappears and only the
normal mode survives.

The next step will be to apply the method presented here to more
realistic cases, namely to the axial and radial breathing modes of a
gas in a cigar-shaped trap containing a larger number of particles. In
fact, the deformation and the large particle number do not pose a big
problem, which is one of the main advantages of the present method as
compared with quantum mechanical QRPA calculations. Another possible
application of the method is to study the dynamics of a vortex, where
already the equilibrium situation is characterized by a non-vanishing
phase of the order parameter.

However, there are still a number of unsolved problems and possible
improvements of the method. First of all, the collision term
\cite{VollhardtWoelfle} should be included, which is an additional
source of damping of the collective oscillations. As mentioned in the
introduction, the possibility to include collisions is an important
advantage of the present method as compared with the QRPA, where
collision effects cannot be taken into account since this would
necessitate to include four-quasiparticle excitations. Second, from a
fundamental point of view, the fact that the continuity equation is
only approximately fulfilled is of course unsatisfactory and one
should think about another numerical method for solving the continuity
equation. Finally, one might ask the question how the present theory
can be extended to the strongly interacting regime. Unfortunately,
this question is up to now completely open, since in this regime
thermal fluctuations of the order parameter, which are not contained
in the BdG equations, play a crucial role (see, e.g., \Ref{Strinati}).
\acknowledgments 
The author wishes to thank P.~Schuck for numerous fruitful discussions
and the critical reading of the manuscript.

\end{document}